\documentclass[journal,review,preprints,moreauthors]{mdpi}

\newcommand{\be}{\begin{equation}}
\newcommand{\ee}{\end{equation}}
\newcommand{\ba}{\begin{array}}
\newcommand{\ea}{\end{array}}
\newcommand{\bea}{\begin{eqnarray}}
\newcommand{\eea}{\end{eqnarray}}

\newcommand{\besub}{\begin{subequations}}
\newcommand{\eesub}{\end{subequations}}

\newcommand{\eg}{e.g.,\ }	
\newcommand{\ie}{i.e.,\ }	

\newcommand{\km}{{\rm km}} 
\newcommand{\cm}{{\rm cm}} 
\newcommand{\s}{{\rm s}} 
\newcommand{\GeV}{{\rm GeV}} 

\newcommand{\kHz}{{\rm kHz}}

\newcommand{\DM}{{\rm DM}} 
\newcommand{\beq}{\begin{equation} \begin{aligned}}
		\newcommand{\eeq}{\end{aligned} \end{equation}}

\definecolor{darkerblue}{rgb}{0.2,0.2,0.5}
\definecolor{seagreen}{rgb}{0.180392,0.545098,0.341176}
\definecolor{smagenta}{rgb}{0.5,0.145098,0.341176}
\definecolor{deepblue}{rgb}{0,0,1}

\firstpage{1} 
\pubvolume{1}
\issuenum{1}
\articlenumber{0}
\pubyear{2023}
\copyrightyear{2023}
\datereceived{ } 
\daterevised{ } 
\dateaccepted{ } 
\datepublished{ } 
\hreflink{https://doi.org/} 
\pdfoutput=1 

\usepackage{caption}
\usepackage{subcaption}

\Title{Solar radio emissions and ultralight dark matter}

\TitleCitation{Title}

\Author{Haipeng An $^{1,2,3,4,*}$\orcidB{}
,
Shuailiang Ge $^{3,5,*}$\orcidG{}
and 
Jia Liu $^{5,3,*}$\orcidA{}
}

\AuthorNames{Firstname Lastname, Firstname Lastname and Firstname Lastname}

\AuthorCitation{Lastname, F.; Lastname, F.; Lastname, F.}

\address{%
$^{1}$ \quad Department of Physics, Tsinghua University, Beijing 100084, China \\
$^{2}$ \quad Center for High Energy Physics, Tsinghua University, Beijing 100084, China\\
$^{3}$ \quad Center for High Energy Physics, Peking University, Beijing 100871, China \\
$^{4}$ \quad Frontier Science Center for Quantum Information, Beijing 100084, China \\
$^{5}$ \quad School of Physics and State Key Laboratory of Nuclear Physics and Technology, Peking University, Beijing 100871, China}

\corres{Correspondence: anhp@mail.tsinghua.edu.cn (H.A.); sge@pku.edu.cn (S.G.); jialiu@pku.edu.cn (J.L.)}

\abstract{
Ultralight axions and dark photons are well-motivated dark matter candidates. Inside the plasma, once the mass of ultralight dark matter candidates equals the plasma frequency, they can resonantly convert into electromagnetic waves, due to the coupling between the ultralight dark matter particles and the standard model photons. The converted electromagnetic waves are monochromatic. In this article, we review the development of using radio detectors to search for ultralight dark matter conversions in the solar corona and solar wind plasma. 
}

\keyword{solar radio emissions; ultralight dark matter; dark photon; axion; resonant conversion}

\begin{document}




\section{Introduction}

There is overwhelming evidence from astronomical and cosmological observations supporting that about one-fourth of the energy density in today's universe is composed of dark matter. However, the particle nature of dark matter is still mysterious. 
Because of the null results of searching for WIMPs~\cite{LUX:2016ggv, XENON:2018voc, PandaX-4T:2021bab}, more and more attention has been attracted to ultralight dark matter candidates such as the QCD axion~\cite{Ipser:1983mw}, axion-like particles~\cite{Svrcek:2006yi}, and dark photons~\cite{Redondo:2008ec, Nelson:2011sf, Arias:2012az, Graham:2015rva}. Originally, the theory of QCD axion was put forward to resolve the strong CP problem via the known Peccei-Quinn mechanism~\cite{Peccei:1977hh, Peccei:1977ur, Weinberg:1977ma, Wilczek:1977pj}. Soon after that, it was shown that, as a bonus, QCD axions are suitable dark matter candidates~\cite{Ipser:1983mw}. In addition, many models such as the string theory will give rise to axion-like particles which have similar features as QCD axions in coupling with the standard model of particle physics. Similarly, axion-like particles can also serve as dark matter candidates. (In the rest of the paper, we will call both the QCD axion and axion-like particles the axions). There are multiple ways in the early Universe of generating  axions: both the misalignment mechanism~\cite{Preskill:1982cy, Abbott:1982af, Dine:1982ah} and the decay of axionic topological defects such as domain walls and strings~\cite{vilenkin1982cosmic,sikivie1982axions} can produce an appropriate amount of axions as dark matter. On the other hand, dark photon is usually a $U(1)$ gauge boson which kinetically mixes with the standard model photons with a tiny coupling strength~\cite{Holdom:1985ag, Dienes:1996zr, Abel:2003ue, Abel:2008ai, Abel:2006qt, Goodsell:2009xc}. Dark photons are also good dark matter candidates~\cite{Redondo:2008ec, Nelson:2011sf, Arias:2012az, Graham:2015rva} and can be generated in multiple ways such as parametric resonances \cite{Co:2018lka, Dror:2018pdh, Bastero-Gil:2018uel, Agrawal:2018vin, Co:2021rhi,  Nakayama:2021avl}, inflationary fluctuations \cite{Graham:2015rva, Ema:2019yrd, Kolb:2020fwh, Salehian:2020asa, Ahmed:2020fhc, Nakai:2020cfw, Nakayama:2020ikz, Kolb:2020fwh, Salehian:2020asa, Firouzjahi:2020whk, Bastero-Gil:2021wsf, Firouzjahi:2021lov, Sato:2022jya}, misalignment mechanism~\cite{Nelson:2011sf, Arias:2012az, AlonsoAlvarez:2019cgw, Nakayama:2019rhg, Nakayama:2020rka}, and cosmic strings decay~\cite{Long:2019lwl}. Current measurements show that the local dark matter density in the solar system, $\rho_{\rm DM}$, is about $0.3~{\rm GeV}/{\rm cm}^3$~\cite{deSalas:2019pee,deSalas:2020hbh}. In this paper, we assume that the dark matter is saturated either by axions or dark photons.

\begin{figure}
    \centering
    \begin{subfigure}{0.9\textwidth}
    \centering
    \includegraphics[width=1\linewidth]{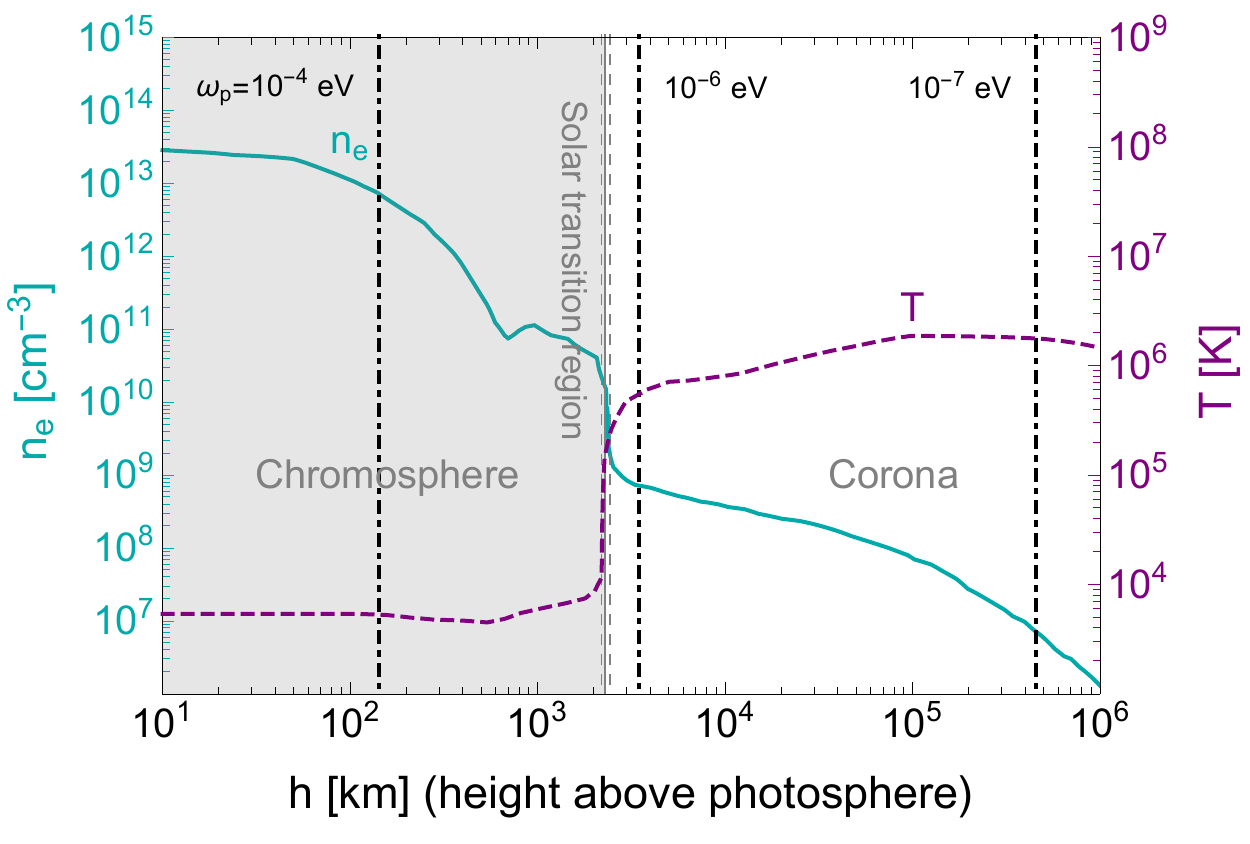}
    \caption{}
    \label{fig:solar_profile_corona}
    \end{subfigure}
    \hfill
    \begin{subfigure}{0.9\textwidth}
    \centering
    \includegraphics[width=1\linewidth]{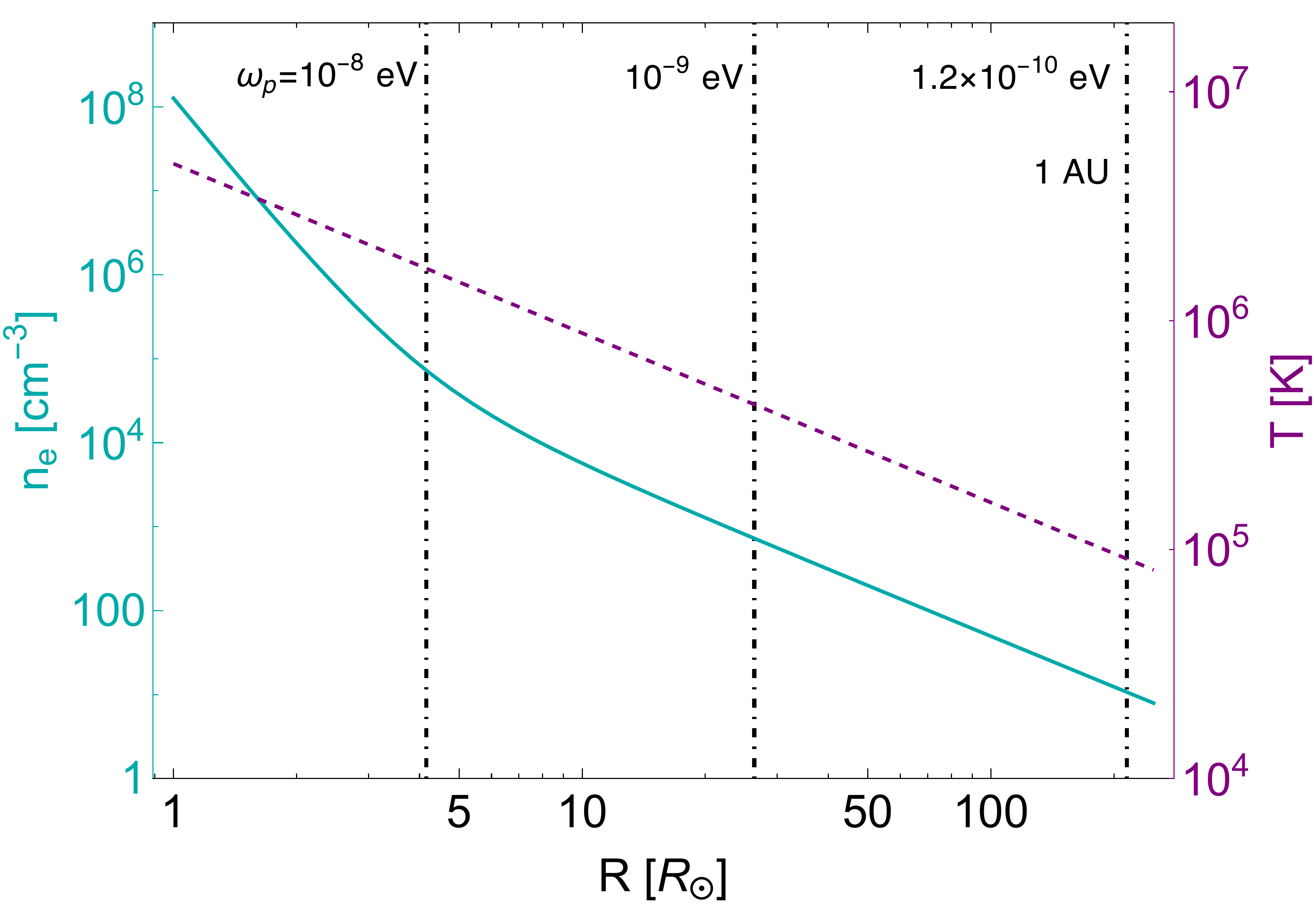}
    \caption{}
    \label{fig:solar_profile_wind}
    \end{subfigure}
 \caption{(a): Quiet-Sun electron density $n_e$ and temperature $T$ profiles in the solar atmosphere above the photosphere. Figure adapted from Ref.~\cite{An:2020jmf} based on the result of  Ref.~\cite{2008GeofI..47..197D}. (b): $n_e$ and $T$ (thermal temperature of electrons) profiles in the solar-wind plasma with the distance extending up to about $1$~AU. The $n_e$ profile is based on the in situ measurements of the Parker Solar Probe~\cite{Moncuquet2020First/FIELDS} and then fitted by the relation $n_e(r)\sim \left[3.3\times10^5 (r/R_{\odot})^{-2} + 4.1\times 10^6 (r/R_{\odot})^{-4} + 8.0\times 10^7(r/R_{\odot})^{-6}\right]~\cm^{-3}$ given in~\cite{1998SoPh..183..165L}. The $T$ profile is the relation $T(r) \sim 418~{\rm eV} \times (r/R_{\odot})^{-0.74}$~\cite{Moncuquet2020First/FIELDS} based on the in situ measurements.  }
 \label{fig:solar_profile}
\end{figure}

The interaction between axions and electromagnetic fields makes it possible for the axion dark matter to convert to electromagnetic waves in a magnetic field. If plasma exists, the interaction between the electromagnetic field and the charged particles in the plasma can induce an effective mass for transverse photon, which equals the plasma frequency $\omega_p$. Thus, in the region that $\omega_p = m_{\rm DM}$, the axion DM will resonantly convert to electromagnetic waves. For dark photon dark matter, the plasma will also regenerate the mixing of the dark photon field and the photon field. Thus, the once diagonalized fields can convert to each other again inside the plasma. Therefore the dark photon dark matter can also convert into electromagnetic waves in the region $\omega_p = m_{\rm DM}$. The energy of the converted photon equals the energy of the dark matter particle, which can be written as $m_{\rm DM} + {E_k}_{\rm DM}$, where ${E_k}_{\rm DM}$ is the kinetic energy of each dark matter particle. In the galactic halo, we have ${E_k}_{\rm DM} \sim 10^{-6} m_{\rm DM}$. Thus, we expect a monochromatic electromagnetic wave signal produced by the ultralight dark matter conversions. To take advantage of the resonant conversion to search for ultralight dark matter, we need first to find a large plasma. Secondly, the region between the conversion region and the detector must be transparent for the converted electromagnetic wave to propagate to the detector. A perfect plasma that we can use is the Sun's corona. The distribution of the temperature and plasma density in the Sun's corona is shown in Fig.~\ref{fig:solar_profile_corona}, where one can calculate that $\omega_p$ ranges from about $10^{-6}$ eV to $10^{-7}$ eV. Thus, if the mass of axion dark matter or dark photon dark matter falls in this region, it can produce monochromatic electromagnetic wave signals. The frequency in this region is about 10 MHz $-$ 1 GHz, which is in the radio wave band. There is already a lot of radio observation apparatus toward the Sun, such as the Low Frequency Array (LOFAR)~\cite{vanHaarlem:2013dsa}, having collected plenty of solar radio data, in which we can search for ultralight dark matter conversion signals. There will also be future projects, such as the Square Kilometer Array (SKA)~\cite{dewdney2009square} and the upgraded Five Hundred Meter Aperture Spherical Radio Telescope (FAST), which may provide more precise solar data in the future.

The earth's ionosphere also provides the plasma for the ultralight dark matter conversion. However, the frequency range is about 4-6 MHz, which is polluted by human activities, and therefore cannot be used to search ultralight dark matter. 
Also, because of the screening effect of the earth's ionosphere, there are no terrestrial radio telescopes with frequency ranges below about 10 MHz. Thus, if the frequency of the ultralight dark matter ($m_{\rm DM}/(2\pi)$) is smaller than 10 MHz, we need to go outer space to search for the conversion signal. The reason the conversion can still happen is that solar wind provides charged particles in the inter-region between the Sun and the Earth. In Fig.~\ref{fig:solar_profile_wind}, one can see that $\omega_p$ in the region between the Sun and the earth is all the way down to $10^{-10}$ eV. Solar probe satellites with long-wave radio spectrometers, such as the STEREO satellites~\cite{kaiser2008stereo} and the Parker Solar Probe~\cite{Pulupa2017TheProcessing}, have been sent out to observe solar activities, and have collected a lot of data. Thus, one can search for the conversion signal in these observation data.

Similar ideas have been proposed to search for radio waves from axion resonant conversion in the mega magnetic field around the neutron stars~\cite{Pshirkov:2007st, Huang:2018lxq, Hook:2018iia} and white dwarfs~\cite{Hardy:2022ufh, Wang:2021wae, Dessert:2019sgw, Dessert:2022yqq}. On the other hand, solar radio emissions are also related to other dark matter models. For instance, the impulse radio events observed by the Murchison Widefield Array (MWA)~\cite{2020ApJ...895L..39M} may be explained by the annihilation of axion-quark-nugget dark matter~\cite{Zhitnitsky:2002qa,Liang:2016tqc,Ge:2017idw,Ge:2017ttc,Ge:2019voa} in the solar corona, the energy released from which may account for the solar corona heating mystery~\cite{Zhitnitsky:2017rop, Raza:2018gpb, Ge:2020xvf}.

In this article, we review the recent developments in searching for ultralight dark matter in solar radio observation. In Sec.~\ref{sec:2}, we review particle physics models for axions and dark photons, focusing on their interactions with the electromagnetic wave. In Sec.~\ref{sec:3}, we present the calculation of the conversion probabilities for axions and dark photons. In Sec.~\ref{sec:4}, we review the calculation of the propagation of the converted electromagnetic waves. We discuss the sensitivities of the radio telescopes for such resonant signals in Sec.~\ref{sec:5}. We summarize the results in Sec.~\ref{sec:6}.






\section{Ultralight dark matter}
\label{sec:2}

Axion is a pseudoscalar particle whose interaction with the electromagnetic part can be written as
\beq \label{eq:lagrangian_axion}
    \mathcal{L}_{\rm axion}=-\frac{1}{4} F^{\mu\nu}F_{\mu \nu} + \frac{1}{2}\partial_{\mu}a\partial^{\mu}a -\frac{1}{2}m_a^2 a^2+\frac{1}{4} a g_{a\gamma\gamma} F^{\mu\nu}\tilde{F}_{\mu \nu}.
\eeq
$a$ is the axion field and $m_a$ is its mass. $F^{\mu\nu} = \partial^{\mu}A^{\nu}-\partial^{\nu}A^{\mu}$ is the field strength of the standard model photon and $\tilde{F}_{\mu \nu} = \varepsilon_{\mu\nu\alpha\beta}/2 \cdot F^{\alpha\beta}$ is the dual field strength. $g_{a\gamma\gamma}$ is the coupling between axion and standard model photon. The interaction term in \eqref{eq:lagrangian_axion} can be further simplified to $-g_{a\gamma\gamma} a \bold{E}\cdot \bold{B}$ in terms of the electric and magnetic fields.

Dark photon is a $U(1)$ gauge boson that, after electroweak symmetry breaking, interacts with the standard model photon via the following Lagrangian:
\beq \label{eq:lagrangian_dp}
    \mathcal{L}_{dp}=-\frac{1}{4}F^{\mu\nu} F_{\mu\nu} -\frac{1}{4}F'^{\mu\nu}F'_{\mu\nu} +\frac{1}{2}m_{A'}^2 A'^2 -\frac{\epsilon}{2}F'^{\mu\nu}F_{\mu\nu}.
\eeq
$F'^{\mu\nu}=\partial^{\mu}A'^{\nu}-\partial^{\nu}A'^{\mu}$ is the field strength of dark photon and $m_{A'}$ is its mass. $\epsilon$ is the kinetic mixing strength between dark photon and photon. To study more conveniently the conversion between dark photon and photon, we can diagonalize \eqref{eq:lagrangian_dp} in the \textit{interaction basis} with the following field redefinition:

\beq
\begin{pmatrix}
A \\
A'
\end{pmatrix}
\rightarrow
\begin{pmatrix}
\frac{1}{\sqrt{1-\epsilon^2}} & 0 \\
 \frac{-\epsilon}{\sqrt{1-\epsilon^2}} & 1 
\end{pmatrix}
\begin{pmatrix}
A \\
A'
\end{pmatrix},
\eeq
and~\eqref{eq:lagrangian_dp} becomes 
\beq \label{eq:lagrangian_dp_interaction-basis}
    \mathcal{L}_{dp}=-\frac{1}{4}F^{\mu\nu} F_{\mu\nu} -\frac{1}{4}F'^{\mu\nu}F'_{\mu\nu} +\frac{1}{2}m_{A'}^2 A'^2 
    -\epsilon m_{A'}^2 A'_{\mu}A^{\mu}.
\eeq

In solar plasma, photons will acquire an effective mass equal to the plasma frequency, $m_{A} = \omega_p$. The plasma is charge neutral consisting of electrons and ions. Ignoring the contributions of ions due to their large masses, we only care about the plasma frequency of electrons, which in the non-relativistic case is 
\beq\label{eq:plasma_frequency}
\omega_p = \sqrt{\frac{4\pi \alpha n_e}{m_e}}.
\eeq
$\alpha$ is the fine structure constant, $m_e$ is the electron mass, and $n_e$ is the electron density. Therefore, for photons in plasma, we should add an effective mass term $\mathcal{L}\supset 1/2 \cdot  \omega_p^2 A^2$ in the Lagrangians~\eqref{eq:lagrangian_axion} and~\eqref{eq:lagrangian_dp_interaction-basis}.

\section{Conversion in solar plasma}
\label{sec:3}
The equations of motions corresponding to the Lagrangians~\eqref{eq:lagrangian_axion} and~\eqref{eq:lagrangian_dp_interaction-basis} for the axion and dark photon cases can both be written in the form~\cite{Raffelt:1987im}
\beq\label{eq:EoM_conversion}
\left(\frac{\partial^2}{\partial t^2} - \frac{\partial^2}{\partial r^2} + \tilde{M} \right)
\begin{pmatrix}
A \\
X
\end{pmatrix}
= 0
,~~~
\tilde{M}=
\begin{pmatrix}
\tilde{\Delta}_{AA} & \tilde{\Delta}_{AX}\\
\tilde{\Delta}_{AX} & \tilde{\Delta}_{XX}
\end{pmatrix}.
\eeq
where $X$ denotes either axion $a$ or dark photon $A'$. For the axion case, we have 
\beq\label{eq:Delta_axion}
\tilde{\Delta}_{AA}= m_A^2 = \omega_p^2,~~~ \tilde{\Delta}_{XX}=m_a^2,~~~ \tilde{\Delta}_{AX}=-g_{a\gamma\gamma} |\boldsymbol{B}_{T}| \cdot \omega,
\eeq
and for the dark photon case, we have
\beq\label{eq:Delta_dp}
\tilde{\Delta}_{AA}= m_A^2 = \omega_p^2,~~~\tilde{\Delta}_{XX}=m_{A'}^2,~~~\tilde{\Delta}_{AX}=-\epsilon m_{A'}^2.
\eeq
Since the external environment in the solar plasma mostly depends on the radial direction $r$ (this is true for the dominant term $\Delta_{AA}$ which is determined by electron density according to ~\eqref{eq:plasma_frequency}), we only keep the spatial derivative in the radial direction in the equation of motion~\eqref{eq:EoM_conversion} while ignoring trivial dynamics in the other two spatial dimensions. 

Some differences between the axion case and the dark photon case are as follows. In comparison with the dark photon case, the conversion between axion and photon requires the presence of magnetic field $\boldsymbol{B}_{T}$ transverse to the axion trajectory. The solar magnetic field provides the source of $\boldsymbol{B}_{T}$. Due to the pseudoscalar feature of axion, the converted photons $A$ in~\eqref{eq:EoM_conversion} is polarized in the direction of $\boldsymbol{B}_{T}$~\cite{Raffelt:1987im,Dessert:2019sgw}. For the dark photon case, \eqref{eq:EoM_conversion} describes converted photons in two transverse modes perpendicular to the propagation trajectory. 

We assume the solutions to the equation of motion~\eqref{eq:EoM_conversion} are in the wave form:
\beq 
A(t,r)=\tilde{A}(r)\exp(-i\omega t+ ik_r r)
,~~~
X(t,r)=\tilde{X}(r)\exp(-i\omega t+ ik_r r)
\eeq
where $\omega$ is the energy of wave, $k=\sqrt{\omega^2-m_X^2}$ is the momentum, and $k_r$ is the momentum component in the radial direction. For resonant conversion to happen, we should have $m_{A}=m_{X}$ such that both $\omega$ and $k$ of $A$ and $X$ match before and after the conversion. In the case we consider here where the dark matter velocity is non-relativistic $v\sim 10^{-3} c$, we have $\omega\simeq m_{X}$. Then, near the resonant layer in solar plasma, we have the following relation~\cite{Raffelt:1987im}:
\beq\label{eq:k_relation}
\frac{\partial^2}{\partial t^2} - \frac{\partial^2}{\partial r^2}
& = -\omega^2 - \frac{\partial^2}{\partial r^2} 
= -\left(k_r+ i\frac{\partial}{\partial r}\right) \left(k_r- i\frac{\partial}{\partial r} \right)-m_{X}^2-k_T^2 
\\
& \simeq -2k_r\left(k_r+ i\frac{\partial}{\partial r}\right) -m_{X}^2-k_T^2.
\eeq
The last approximated relation holds because the parameters of solar plasma, \eg plasma frequency, vary slowly in a length scale much larger than the wavelength $1/k_r$ such that
\beq
\left| \frac{\partial}{\partial r}\tilde{A}(r) \right|\ll k_r |\tilde{A}(r) |
,~~~
\left| \frac{\partial^2}{\partial r^2}\tilde{A}(r) \right|\ll k_r \left| \frac{\partial}{\partial r}\tilde{A}(r) \right| .
\eeq
Therefore, using the relation~\eqref{eq:k_relation}, the equation of motion~\eqref{eq:EoM_conversion} can be simplified to
\beq\label{eq:EoM_conversion_simplified}
\left(i\frac{\partial}{\partial r} - M \right) 
\begin{pmatrix}
\tilde{A}(r) \\
\tilde{X}(r)
\end{pmatrix}
= 0
\eeq
where 
\beq
& M = \frac{1}{2k_r} (\tilde{M} - m_{X}^2) 
\equiv  M_0 + M_1,
\\
& M_0 =  \begin{pmatrix}
\frac{1}{2k_r}(\omega_p^2- m_{X}^2 -k_T^2) & 0\\
0 & - \frac{1}{2k_r} k_T^2
\end{pmatrix}
,~~~
M_1=\begin{pmatrix}
0 & \Delta_{AX}\\
\Delta_{AX} & 0
\end{pmatrix}
,~~~
\Delta_{AX} \equiv \frac{ \tilde{\Delta}_{AX} }{2k_r}.
\eeq

The off-diagonal matrix $M_1$ is suppressed by the coupling constant, either $\epsilon$ or $g_{a\gamma\gamma}$, and thus can be treated as a perturbative term. If we start with a $X$ particle, $(A, X) = (0,1)$ initially, then solving the equation of motion~\eqref{eq:EoM_conversion_simplified} to first order of $M_1$, the probability that $X$ eventually converts to a photon $A$, namely, $(A, X) = (1,0)$, is 
\beq\label{eq:P_pert-method}
P_{X\rightarrow \gamma} = \left|
\int_{r_0}^r dr' 
\Delta_{AX}(r')
\exp{\left\{i\int_{r_0}^{r'} dr''\frac{1}{2k_r} \left[\omega_p^2(r'')- m_{X}^2 \right]\right\}}
\right|^2.
\eeq
We can solve the integration using the saddle-point method
\beq\label{eq:saddle-point-appro}
\left|\int dr'~ g(r') {\rm e}^{-ih(r')} \right| \approx g(r_c)\sqrt{\frac{2\pi}{|h''(r_c)|}}
,~~~
h(r') \equiv \int_{r_0}^{r'} dr''\frac{1}{2k_r} \left[\omega_p^2(r'')- m_{X}^2 \right]
\eeq
where we have used the expression of norm to remove the redundant phase factor. $r_c$ is the position when $h(r)$ reaches its local maximum/minimum $h'(r_c)=0$, that is, the position of resonant $X$-$A$ conversion when $\omega_p(r_c) = m_{X}$. Only the second derivative plays an important role, the expression of which is 
\beq
h''(r_c) = 
\frac{\omega_p}{k_r} \left. \frac{\partial \omega_p(r)}{\partial r} \right|_{r=r_c} .
\eeq
The length of resonant layer $\delta r_c$ can be estimated as the length that makes $h(r)\approx h(r_c)+1/2 \cdot (r-r_c)^2 h''(r_c)$ change by $2\pi$, so we get 
\beq
\delta r_c
\sim \sqrt{\frac{4\pi}{|h''(r_c)|}} 
\sim \sqrt{\frac{r_c v_r}{\omega_p}}.
\eeq
The resonant layer is very thin, $\delta r_c/r_c \sim 10^{-6}$, for the typical values $r_c = R_{\odot}$, $\omega_p = 2\pi \cdot 100$~MHz, and $v_r = 10^{-3}c$.

Applying the saddle-point approximation~\eqref{eq:saddle-point-appro}, we solve \eqref{eq:P_pert-method} and obtain
\beq
P_{X\rightarrow \gamma} 
= \Delta_{AX}^2(r_c)\cdot  \frac{2\pi k_r}{\omega_p}\left|\frac{\partial\omega_p}{\partial r}\right|^{-1}_{r=r_c} .
\eeq
Then, for each case we have
\beq
	P_{a\rightarrow \gamma} = \pi \frac{g_{a\gamma\gamma}^2 \left|\bold{B}_{T}\right|^2 }{m_a}
	\frac{1}{v_r(r_c)}  \left |\frac{\partial \ln \omega_p^2}{\partial r} \right |^{-1}_{r=r_c}
 ,~~~
 \text{(axion case)}
\eeq
and
\beq
P_{A'\rightarrow \gamma}
\simeq
\frac{2}{3} \pi \epsilon^2 m_{A'} \frac{1}{v_r(r_c)}  \left|\frac{ \partial \ln\omega_p^2}{\partial r}\right|^{-1}_{r=r_c}
,~~~
\text{(dark photon case)} .
\eeq
Note that we have an extra prefactor $2/3$ in the dark photon case. This is because the converted photons polarize in three directions with equal probability but the longitudinal mode cannot propagate out of the plasma. In comparison, the photons converted from axions polarize only in one transverse mode parallel to $\boldsymbol{B}_{T}$ as mentioned above. Furthermore, the conversion probabilities for the two cases are related via the following replacement
\beq\label{eq:axion_DP_equiv}
\sqrt{\frac{2}{3}}\epsilon m^2_{A'} \Leftrightarrow  g_{a\gamma\gamma} \left|\bold{B}_{T}\right| \omega
,~~~
\text{($\omega\simeq m_a$, non-relativistic).}
\eeq
This relation is useful when we know the conversion rate of one case and directly calculate that of the other case.


\section{Propagation of the converted photons}
\label{sec:4}
After the radio photons are converted from axion or dark photon, they will be refracted, absorbed, and scattered by thermal electrons in the plasma. We discuss these effects during the photon propagation in the following.

\noindent \textbf{Refraction effect}-- The behavior of converted photons propagating in the solar plasma obeys the rule of refraction:
\beq\label{eq:refraction}
\frac{n(r_c)}{n(r)} = \frac{\sin\theta(r)}{\sin\theta(r_c)}.
\eeq
$\theta(r)$ is the angle of incidence or refraction with respect to the radial direction. $n(r)$ is the refraction index as a function of distance, which is equal to the inverse of phase velocity:
\beq
n(r) = \frac{k}{\omega} = \sqrt{1-\frac{\omega_p^2(r)}{\omega^2}}.
\eeq
At the position of resonant conversion, $n(r_c)\simeq v \sim 10^{-3}$. Once the converted photons leave the resonant layer, the refraction index $n(r)$ quickly increases to values close to $1$ as the number density of electrons $n_e$ (and thus the plasma frequency, according to \eqref{eq:plasma_frequency}) abruptly decreases with the distance. For example, as we can see from Fig.~\ref{fig:solar_profile_corona}, $n_e$ drops by about two orders of magnitude from the height $10^4$~km to the height $10^6$~km. Therefore, the converted photons will quickly propagate nearly radially $\theta\simeq \pi/2$ after traveling a distance based on the refraction law~\eqref{eq:refraction}. 
\\

\noindent \textbf{Absorption effect}-- In addition to the refraction effect discussed above, the converted photons could be absorbed by interacting with the plasma along the propagation. The absorption effect is mainly due to the inverse bremsstrahlung process, the rate of which is estimated as~\cite{An:2020jmf,Redondo:2008aa}
\beq\label{eq:inverse_B}
\Gamma_{\rm invB} \simeq \frac{8\pi \alpha^3 n_e n_I}{3 m_e^2 \omega^3} \sqrt{\frac{2\pi m_e}{T}} \ln\left( \frac{2T^2}{\omega_p^2} \right) \left(1-{\rm e}^{-\omega/T}\right).
\eeq
$n_I$ is the number density of ions in the plasma. This equation is derived under the condition $\omega\ll T \ll m_e$, which is satisfied in both solar corona and solar wind plasma, as can be seen in Fig.~\ref{fig:solar_profile}. Taking into consideration the long-range Coulomb force cut off by the plasma screening effect (Debye screening length $\lambda_d\sim \sqrt{T/(m_e \omega_p^2)}$), we have the $\ln$ factor in \eqref{eq:inverse_B}. At first glance, it seems that the inverse bremsstrahlung rate $\sim \omega^{-3}$ will diverge for low-energy photons, but the numerator factor $n_e n_I \simeq n_e^2 \propto \omega_p^4 \propto \omega^4$ at the resonant layer actually makes $\Gamma_{\rm invB}$ decease quickly as $\omega\rightarrow 0$. 
\\

\noindent  \textbf{Scattering effect}-- The converted photons also interact with plasma via Compton scattering, the rate of which is
\beq\label{eq:Compton}
\Gamma_{\rm Com} = \frac{8\pi \alpha^2}{3m_e^2}n_e.
\eeq
Such interaction can shift the photon energy and propagation direction. We will elaborate more on this effect in the following.

To estimate how the converted photons are affected by the effects of absorption and scattering, one can define an attenuation factor by simply adding the two effects together, $\Gamma_{\rm att} = \Gamma_{\rm invB} + \Gamma_{\rm Com}$~\cite{An:2020jmf}. Then, the survival probability of photons after conversion can be estimated as~\cite{An:2020jmf}
\beq\label{eq:prop_sur_0}
P_{\rm sur} \simeq \exp{\left(-\int_{r_c}^{\infty}[\Gamma_{\rm att}(r)]/v_r dr\right)}.
\eeq
It denotes the probability that a converted photon is neither absorbed nor scattered during propagation. 

Compared with the inverse bremsstrahlung process~\eqref{eq:inverse_B}, the Compton scattering plays a minor role. This can be seen by taking the ratio of~\eqref{eq:Compton} to~\eqref{eq:inverse_B} which gives $\sim (m_e/T)^{-3/2}\ll 1$. In addition, the propagating photons can scatter with random electron density fluctuations (see \eg Ref.~\cite{Kontar_2019}), which is labeled as `irregular refractions' in Ref.~\cite{2007ApJ...671..894T}. Ref.~\cite{Kontar_2019} treated the density fluctuations effectively static, and therefore the scatterings do not change the photon energy but only the photon propagating direction.

Therefore, \eqref{eq:prop_sur_0} represents a simplified and conservative scenario. More realistically, the converted photons will be scattered into different directions and get an angular distribution. As pointed out in Ref.~\cite{Kontar_2019}, the directions of solar radio emissions will be quickly randomized by scattering effect (mainly due to the electron density fluctuations) near the resonant region where the plasma frequency is equal to the emission frequency, while the large-scale refraction generally tends to shift the propagation path towards the radial direction. Solving exactly the evolution dominated by repeated scatterings and inverse bremsstrahlung processes is difficult, and this problem is usually transformed into the framework of Fokker-Planck equation (see \eg Refs.~\cite{Kontar_2019,Bian_2019,acharya2021comparison,cooper1971compton}). Essentially, the Fokker-Planck equation can describe the evolution of the phase-space distribution $N(\boldsymbol{r}, \boldsymbol{k}, t)$ of photons, where the refraction effect represented by the gradually-varying $n_e(r)$, the absorption effect represented by a collision absorption coefficient, and the scattering effect represented by a diffusion tensor, are included~\cite{Kontar_2019,Bian_2019}. Based on the Fokker-Planck equation, Ref.~\cite{Kontar_2019} has developed a numerical Monte Carlo ray-tracing method to study the 3-dimension stochastic propagation process of radio emissions where the anisotropic $n_e$ fluctuation has been considered. 

Applying the ray-tracing numerical code~\cite{Kontar_2019}, Ref.~\cite{An:2023wij} has studied the propagation of converted photons in the plasma. The survival probability $P_{\rm sur}$ defined in the numerical study is only related to absorption regardless of whether scattered or not, which is different from that in~\eqref{eq:prop_sur_0}. The scattering effect against the refraction effect is represented by the angular distribution function $g(\theta)$ where $\theta$ is the angle between the final propagation direction and the radial direction on the last scattering surface ($\sim 6 R_{\odot}$, beyond which the interactions of photons with the plasma are negligible). The numerical results show that $P_{\rm sur}\simeq 0.5-0.07$ and $g(\theta)$ distributes mostly within the range $\theta\sim 0.15~{\rm rad}$, in the radio frequency range $30-80$~MHz, and both the two effects are stronger for higher frequencies.

\section{Detection}
\label{sec:5}

For dark matter particles with typical initial velocity $v_0\simeq 220 ~ \km/\s$, the solar gravitational effect cannot be neglected when they travel close to the Sun since $GM_{\odot}/(R_{\odot} v_0^2)\simeq 3.45$ is an $\mathcal{O}(1)$ number. The trajectory of a dark matter particle from far away is hyperbolic in the solar gravitational potential. In the polar coordinates $(r,\theta)$ centered at the Sun, the dark matter trajectory can be expressed as 
\beq\label{eq:trajectory}
r = \frac{v_0^2 b^2}{G M_{\odot}}\frac{1}{1+e\cos\theta}
,~~~
e = \sqrt{1+b^2\left(\frac{v_0^2}{GM_{\odot}}\right)^2}
\eeq
where $b$ is the impact factor and $e$ is the eccentricity of trajectory. The tangential velocity $v_{\theta}$ and radial velocity $v_r$ of dark matter are respectively
\beq
v_{\theta} = v_0 \frac{b}{r}
,~~~
v_{r} = \sqrt{v_0^2- v_{\theta} ^2 +\frac{2GM_{\odot}}{r}}
,~~~
v^2 = v_r^2 + v_{\theta}^2.
\eeq

The dark matter velocity $v_0$ follows the Maxwellian distribution in the Galactic frame~\cite{Drukier:1986tm,Choi:2013eda,Evans:2018bqy}:
\beq\label{eq:Maxwellian-Galaxy}
f_G(v_0) = \frac{4}{\sqrt{\pi}} \frac{v_0^2}{v_p^3}\exp\left(-\frac{v_0^2}{v_p^2}\right),
\eeq
and we have $\int_{0}^{\infty}d v_0 f_G(v_0)=1$. $v_p$ is the most probable speed, which is taken as the speed of Local Standard of Rest, \ie the circular speed about the Galactic center at the solar position, $v_p\simeq v_{\odot}\simeq 220 ~\km/\s$. The distribution \eqref{eq:Maxwellian-Galaxy} should be truncated at $v_{\rm esp}$ which is the Galactic escape speed at the solar position, $v_{\rm esp} \simeq 544~\km/\s$. However, this modification is exponentially tiny as $\int_{v_{\rm esp}}^{\infty}d v_0 f_G(v_0)\approx 0.66\%$,
which can be safely ignored. Next, we make a Galilean boost to get the velocity distribution in the rest frame of the Sun (see \eg Ref.~\cite{Choi:2013eda}):
\beq\label{eq:Maxwellian_Sun_frame}
&f(v_0)=\frac{1}{\sqrt{\pi}}\frac{v_0}{v_p v_{\odot}}
\left\{
\exp\left[-\frac{(v_0-v_{\odot})^2}{v_p^2}\right] - 
\exp\left[-\frac{(v_0+v_{\odot})^2}{v_p^2}\right]
\right\} .
\eeq

Dark photons or axions in the dark matter wind can reach the resonant-conversion layer $r_c$ if their impact parameters $b$ 
are within
\beq
b_{\rm max} = r_c \frac{v(r_c)}{v_0} =r_c\sqrt{1+\frac{2GM_{\odot}}{v_0^2 r_c}}.
\eeq
Then, the total power of the converted photons per unit of solid angle flowing out of the sphere $r_c$ is~\cite{An:2020jmf}
\beq\label{eq:Power_0}
\frac{d\mathcal{P}_0}{d\Omega} = 2\times \frac{1}{4\pi} \rho_{\rm DM} v_0 \cdot \int_{0}^{b_{\rm max}}db 2\pi b \cdot P_{X\rightarrow \gamma}
= r_c^2  P_{X\rightarrow \gamma}(v_0) \rho_{\DM} v(r_c). 
\eeq
The prefactor $2$ accounts for the fact that the dark matter particles entering and leaving the resonant sphere can both resonantly convert into photons with equal probability. The converted photons propagating inward will be totally reflected as the plasma frequency is larger in deeper region. Then, with the Maxwellian velocity distribution~\eqref{eq:Maxwellian_Sun_frame} taken into consideration, the emitted power~\eqref{eq:Power_0} is corrected as~\cite{An:2023wij}
\beq\label{eq:Power}
\frac{d\mathcal{P}}{d\Omega} = \int_{0}^{\infty} dv_0 f(v_0) \frac{d\mathcal{P}_0}{d\Omega}. 
\eeq

The converted photons will generate a spike-like feature superposed on the normal solar radio spectrum. Such signals can be detected by the radio telescopes such as SKA and LOFAR when they reach the Earth. For Earth-based radio telescopes, the spectral flux density (in unit of ${\rm W}/{\rm m}^2/{\rm Hz}$) received is thus~\cite{An:2020jmf,An:2023wij} 
\beq\label{eq:S_sig}
S_{\rm sig} = \frac{1}{d^2} \frac{1}{\mathcal{B}} \frac{d\mathcal{P}}{d\Omega} \cdot P_{\rm sur} \cdot \beta[g(\theta)].
\eeq
The factor $\beta$ represents the suppression on the final detected signal due to the scattering effect. It depends on the angular distribution $g(\theta)$ that requires a detailed ray-tracing numerical simulation~\cite{An:2023wij}. If there is no scattering effect, then $\beta = 1$. $\mathcal{B}$ is the frequency bandwidth over which the signal distributes. We average the signal over the bandwidth and get the signal per unit of frequency as shown in~\eqref{eq:S_sig}. $\mathcal{B}$ should be chosen as the larger one between the telescope resolution $\mathcal{B}_{\rm res}$ and the natural bandwidth $\mathcal{B}_{\DM}$. Due to the velocity dispersion of dark matter, the converted photons have a natural bandwidth of frequency~\cite{An:2020jmf,An:2022hhb}:
\beq\label{eq:bandwidth_DM}
\mathcal{B}_{\DM} \simeq \frac{1}{2\pi} m_{X}v_p^2 \simeq 0.15~\kHz \times \left(\frac{m_X}{10^{-6}~{\rm eV}}\right).
\eeq

On the other hand, we need to know the telescope's ability to detect weak signals. The minimal detectable spectral flux density of a telescope is~\cite{SKA1-Baseline,An:2020jmf}
\beq\label{eq:S_min}
S_{\rm min}= \frac{{\rm SEFD}}{\eta_s \sqrt{n_{\rm pos} \mathcal{B} t_{\rm obs}}}.
\eeq
$\eta_s$ is the system efficiency of a telescope, which is 0.9 and 1 for SKA and LOFAR respectively~\cite{SKA1-Baseline,Nijboer:2013dxa}; $n_{\rm pol} = 2$ is the number of polarizations of photons; $t_{\rm obs}$ is the observation time. SEFD is the system equivalent (spectral) flux density:
\beq
{\rm SEFD} =2 k_B \frac{T^{\rm sys}+T^{\rm nos}_{\odot}}{A_{\rm eff}}.
\eeq
$T^{\rm sys}$ is the system temperature of the radio telescope antenna; $T^{\rm nos}_{\odot}$ is the noise temperature induced on the antenna when pointing towards the Sun, which can be estimated as the black-body temperature of the quiet Sun~\cite{An:2020jmf}; $A_{\rm eff}$ is the effective area of the radio telescope array such as SKA and LOFAR.

Then, taking the signal~\eqref{eq:S_sig} equal the telescope detection ability~\eqref{eq:S_min}, 
\beq\label{eq:S_equal}
S_{\rm sig}=S_{\rm min},
\eeq
one can find the upper limit of the coupling $g_{a\gamma\gamma}$ for the axion case or $\epsilon$ for the dark photon case that can be detected by the telescope. Eq.~\eqref{eq:S_equal} reflects how large the parameter space of dark photon/axion dark matter can be explored via observing solar radio emissions. 

Ref.~\cite{An:2020jmf} has used the SKA phase 1 (SKA1) and LOFAR as benchmark telescopes. To be specific, the following frequency ranges have been considered: the low-frequency band, SKA1-Low $(50-350)$~MHz; two middle-frequency bands, SKA1-Mid B1 $(350-1050)$~MHz and B2 $(950-1760)$~MHz; and two LOFAR frequency bands, $(10-80)$~MHz (low-band antenna) and $(120-240)$~MHz (high-band antenna). Note that LOFAR does not operate in the gap frequency range $(80-120)$~MHz due to the contamination from the man-made frequency-modulation (FM) broadcast band that spans over $\sim(87-108)$~MHz~\cite{vanHaarlem:2013dsa,2009IEEEP..97.1431D}.

The effective area $A_{\rm eff}$ for SKA1 array is about $10^4-10^5~{\rm m}^2$, while it is about $10^3~{\rm m}^2$ for LOFAR. The antenna system temperatures $T^{\rm sys}$ are also given in Ref.~\cite{An:2020jmf} based on the data of Ref.~\cite{SKA1-Baseline, vanHaarlem:2013dsa}, ranging from $\sim 10$ to $10^4$~K for the five frequency bands. Lastly, the telescope resolution $\mathcal{B}_{\rm res}$ for SKA1-Low, SKA1-Mid and LOFAR are respectively 1, 3.5 and 197~kHz. All of them are larger than the natural bandwidth of dark matter~\eqref{eq:bandwidth_DM} for the radio frequency bands we are interested in. Therefore, one should choose $\mathcal{B}=\mathcal{B}_{\rm res}$ when computing the received signal~\eqref{eq:S_sig}.

\begin{figure}
    \centering
    \includegraphics[width=0.75\linewidth]{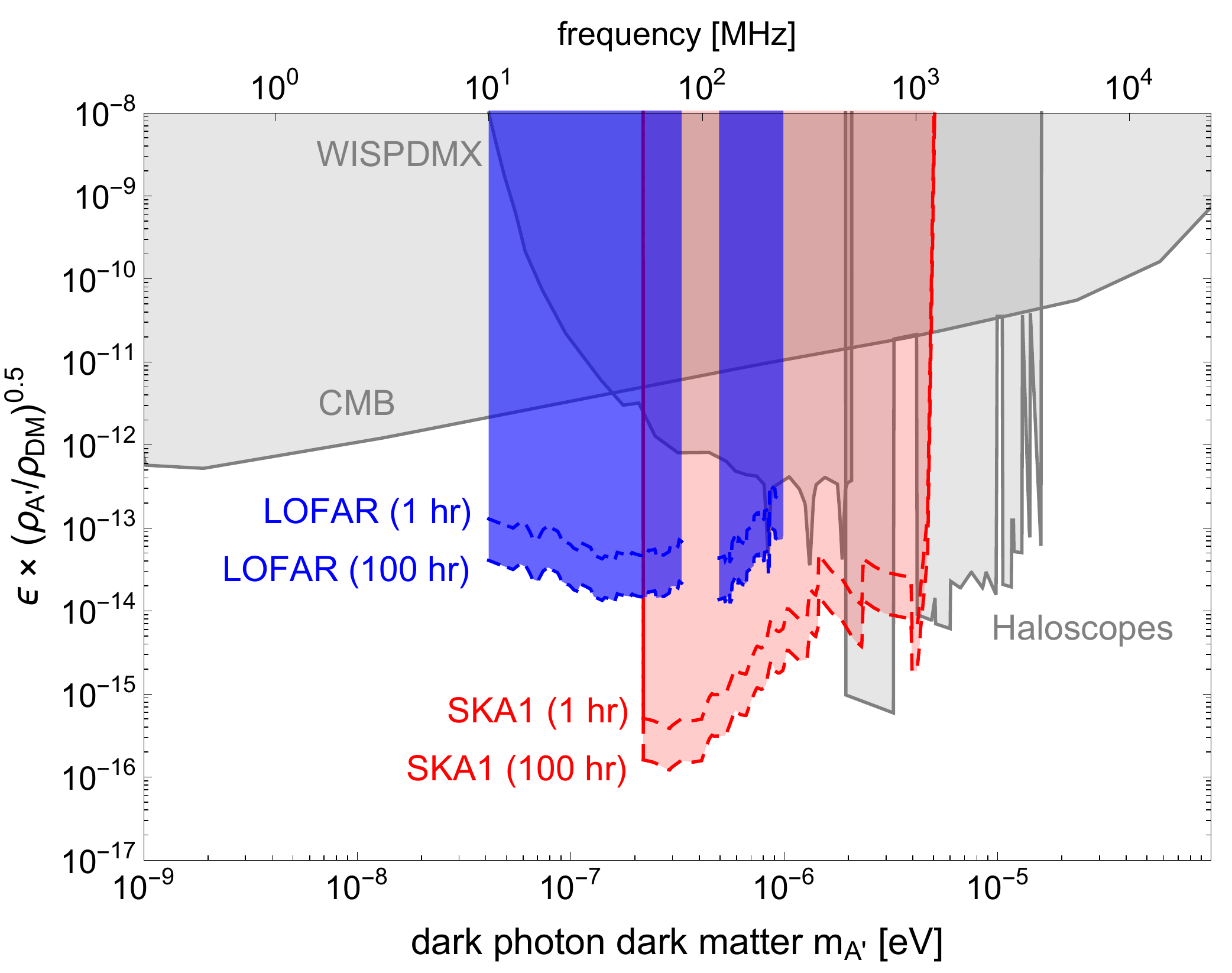}
    \caption{The projected sensitivity reaches of the parameter space of the dark photon dark matter based on the radio telescopes SKA phase 1 and LOFAR, with the observation time assumed to be 1 hour and 100 hours respectively. The plot also shows the existing constraints from CMB~\cite{Arias:2012az, McDermott:2019lch}, the cavity experiment WISPDMX~\cite{Nguyen:2019xuh}, and the multiple haloscope-type experiments~\cite{Arias:2012az, DePanfilis:1987dk, Wuensch:1989sa, Hagmann:1990tj, Asztalos:2001tf, Asztalos:2009yp}. 
    Figure adapted from Ref.~\cite{An:2020jmf}.
    }
    \label{fig:sensitivity_reach}
\end{figure}

With the above properties of the radio telescopes known, the sensitivity reaches for the dark photon case based on \eqref{eq:S_equal} is plotted in Fig.~\ref{fig:sensitivity_reach}. The observation time is assumed to be 1 hour and 100 hours respectively. As shown in the plot, the sensitivity reaches of the kinetic mixing parameter $\epsilon$ based on SKA and LOFAR radio telescopes could be several orders of magnitude better than the existing constraints from the cosmic microwave background (CMB)~\cite{Arias:2012az, McDermott:2019lch} and the cavity experiment WISPDMX~\cite{Nguyen:2019xuh}.

\begin{figure}
    \centering
    \includegraphics[width=0.75\linewidth]{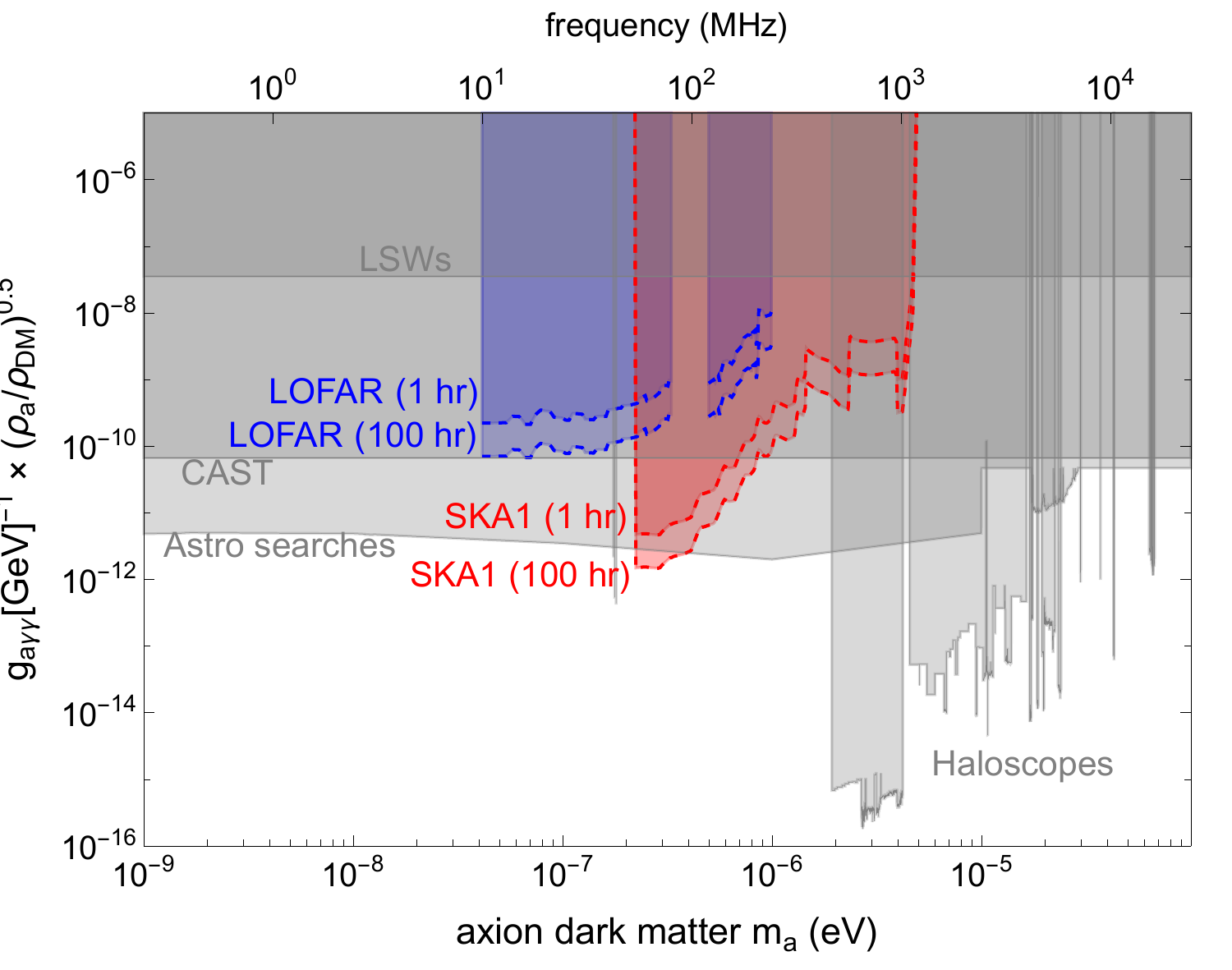}
    \caption{The projected sensitivity reaches of the parameter space of the axion photon dark matter based on the radio telescopes SKA phase 1 and LOFAR, with the observation time assumed to be 1 hour and 100 hours respectively. The results are translated from Fig.~\ref{fig:sensitivity_reach} via the relation~\eqref{eq:axion_DP_equiv} with $|\boldsymbol{B}_T|$ taken as $1$ Gauss~\cite{yang2020global}. Together we show the existing constraints, which can be classified into four categories: the astrophysical searches for axions in various environments, including white dwarfs, neutron stars and pulsars, quasars and blazars, the radio galaxy NGC 1275, and Globular Clusters, etc.~\cite{Dessert:2022yqq,Ayala:2014pea,Dolan:2022kul,Noordhuis:2022ljw,Li:2020pcn,Li:2021gxs, Davies:2022wvj, Fermi-LAT:2016nkz}; various haloscope experiments aiming to detect axions in the Galactic dark matter halo~\cite{DePanfilis:1987dk,Hagmann:1990tj, Asztalos:2009yp, ADMX:2018gho, 0004-637X-490-2-493, ADMX:2019uok, ADMX:2018ogs, HAYSTAC:2018rwy, HAYSTAC:2020kwv, Alesini:2020vny, Lee:2020cfj, Jeong:2020cwz, CAPP:2020utb, ADMX:2021nhd, Crisosto:2019fcj, Lee:2022mnc, Kim:2022hmg, Yi:2022fmn, Adair:2022rtw, HAYSTAC:2023cam, Quiskamp:2022pks, Alesini:2019ajt, Alesini:2022lnp, TASEH:2022vvu}; the helioscope, CAST, aiming to detect axions emitted by the Sun; and Light-Shining-through-a-Wall (LSW) experiments including CROWS, ALPS, and OSQAR~\cite{Betz:2013dza, Ehret:2010mh, OSQAR:2015qdv}. The existing constraints are also nicely summarized in Ref.~\cite{AxionLimits}.  
}
\label{fig:sensitivity_reach_axion}
\end{figure}

The sensitivity reaches obtained in Fig.~\ref{fig:sensitivity_reach} can be translated to the axion case via the relation~\eqref{eq:axion_DP_equiv}, and the result is plotted in Fig.~\ref{fig:sensitivity_reach_axion}. As shown in Ref.~\cite{yang2020global}, the magnetic field strength is about $1-4$ Gauss in the solar corona at $1.05-1.35 R_{\odot}$ from the Sun center. To be conservative, we take the transverse magnetic field $|\boldsymbol{B}_T|=1$~Gauss in~\eqref{eq:axion_DP_equiv}. As we can see from Fig.~\ref{fig:sensitivity_reach_axion}, our projected sensitivities are better than the Light-Shining-through-a-Wall experiments (LSW, including CROWS, ALPS, and OSQAR~\cite{Betz:2013dza, Ehret:2010mh, OSQAR:2015qdv}). Also, they are better than the constraint from the CAST experiment in part of the SKA1 frequency range. However, in most of the relevant frequency range, they cannot exceed the constraints from multiple astrophysical searches in various environments~\cite{Dessert:2022yqq,Ayala:2014pea,Dolan:2022kul,Noordhuis:2022ljw,Li:2020pcn,Li:2021gxs, Davies:2022wvj, Fermi-LAT:2016nkz} (which is shown as a combined constraint in Fig.~\ref{fig:sensitivity_reach_axion}). The performance of sensitivity reaches in the axion case is not as good as that in the dark photon case. This is because the solar magnetic field is relatively low which suppresses the axion-photon conversion rate as we can see from~\eqref{eq:axion_DP_equiv}.

Figs.~\ref{fig:sensitivity_reach} and~\ref{fig:sensitivity_reach_axion} are based on the simplified scenario of the propagation process of the converted photons that took $P_{\rm sur}$ as \eqref{eq:prop_sur_0} and assumed $\beta=1$ in~\eqref{eq:S_sig}. Instead, Ref.~\cite{An:2023wij} has numerically simulated the radio photon propagation based on the Monte Carlo ray-tracing method. Also, the Maxwellian velocity distribution has been considered in computing the emitted power~\eqref{eq:Power}, although this effect does not make much difference. Furthermore, Ref.~\cite{An:2023wij} has analyzed real data of LOFAR (three samples of 17 min observations at different dates). Finally, Ref.~\cite{An:2023wij} gets a constraint up to $\epsilon\sim 10^{-13}$ which is about 1 to 2 orders of magnitude better than the CMB constraint~\cite{Arias:2012az, McDermott:2019lch} in the frequency range $30-80$~MHz. This constraint is weaker than the projections in Fig.~\ref{fig:sensitivity_reach}, mainly because the angular distribution in~\eqref{eq:S_sig} of converted photons based on the numerical simulation has been considered. A similar constraint on the axion-photon coupling strength $g_{a\gamma\gamma} \sim 10^{-9}~\GeV^{-1}$ has been obtained in the same frequency range~\cite{An:2023wij}, which is about 1 to 2 orders of magnitude better than the existing constraints from the LSW experiments~\cite{Betz:2013dza,Ehret:2010mh, OSQAR:2015qdv}, but does not exceed the constraints from the CAST experiment~\cite{CAST:2017uph} and multiple astrophysical bounds~\cite{Dessert:2022yqq,Ayala:2014pea,Dolan:2022kul,Noordhuis:2022ljw,Li:2020pcn,Li:2021gxs, Davies:2022wvj, Fermi-LAT:2016nkz}. 

In addition to the terrestrial facilities such as SKA and LOFAR discussed above, space radio probes are also powerful tools for searching for ultralight dark matter signals from resonant conversion. The STEREO observatory~\cite{kaiser2008stereo, zaslavsky2011antenna} consists of two nearly identical satellites, observing the Sun in the 1 AU orbit. It carries a low-frequency radio receiver operating in the frequency range 2.6 kHz $-$ 150 kHz and a higher-frequency receiver operating in the frequency range 125 kHz $-$ 16 MHz. Furthermore, the Parker Solar Probe~\cite{Pulupa2017TheProcessing,Moncuquet2020First/FIELDS} provides new opportunities for observing the conversion signal
with the perihelion reaching $\sim 10 R_{\odot}$ from the Sun center. In fact, the Parker satellite is the first spacecraft that can fly into the solar corona.  The best advantage for us is that Parker satellite can measure in situ the resonant-conversion events in the solar-wind plasma as shown in Fig.~\ref{fig:solar_profile_wind}, thanks to its motion between the Sun and 1 AU. The Parker satellite carries a low-frequency radio receiver 
operating in the frequency range 10 kHz $-$ 1.7 MHz and a higher-frequency radio receiver operating in the frequency range 1.3 MHz $-$ 19.2 MHz. We expect that the Parker satellite has a strong performance in searching for resonant-conversion signals in these frequency ranges. The analysis of space probes such as Paker Solar Probe and STEREO to search for ultralight dark matter is currently in progress by the authors of this Review.

\section{Summary}
\label{sec:6}
Solar plasma is an ideal place for detecting ultralight axions or dark photons. When axion dark matter or dark photon dark matter particles cross the solar corona or solar wind, they can resonantly convert into standard model photons when their mass equals the plasma frequency. The plasma frequency of solar corona and solar wind spans from 1~GHz to 0.1~MHz 
for the distance extending up to 1 AU. Therefore, searching for the radio signals converted from ultralight axions or dark photons in the plasma provides a complementary method to other haloscope studies, to detect these dark matter particles in the mass range $\sim 10^{-6} - 10^{-10}$ eV and set stringent constraints on the parameter space of these dark matter models. Based on this mechanism,  radio telescopes such as LOFAR and SKA perform well in searching for ultralight dark matter, especially for the dark photon case. 
Future projected SKA and upgraded FAST are expected to obtain solar radio data with much higher precision. 
In addition to the terrestrial telescopes, we can use space solar probes such as STEREO and Parker to search for the ultralight dark matter. We conclude that searching for the resonant-conversion radio signals from solar plasma is an efficient method, which is competitive and complementary compared to other direct detection experiments~\cite{Nguyen:2019xuh,Arias:2012az, DePanfilis:1987dk, Wuensch:1989sa, Hagmann:1990tj, Asztalos:2001tf, Asztalos:2009yp,Betz:2013dza,Ehret:2010mh, OSQAR:2015qdv,CAST:2017uph} aiming for the ultralight dark matter.




\vspace{6pt} 




\funding{The work of H.A. is supported in part by the National Key R$\&$D Program of China under Grant No. 2021YFC2203100 and 2017YFA0402204, the NSFC under Grant No. 11975134, and the Tsinghua University Initiative Scientific Research Program. The work of SG is supported by NSFC under Grant No. 12247147, the International Postdoctoral Exchange Fellowship Program, and the Boya Postdoctoral Fellowship of Peking University. The work of J.L. is supported by NSFC under Grant No. 12075005, 12235001.}

\appendixtitles{yes} 



\begin{adjustwidth}{-\extralength}{0cm}

\reftitle{References}



\bibliography{references}

\PublishersNote{}
\end{adjustwidth}
\end{document}